\documentclass[a4paper]{jpconf}
\usepackage{graphicx}

\begin{document}

\title{Spherical  neutron polarimetry study of hysteresis in multiferroic MnWO$_4$}

\author{A Poole$^1$, A S Wills$^1$, P J Brown$^2$}

\address{$^1$Department of Chemistry, University College London,
20 Gordon Street, London, WC1H 0AJ, UK}
\address{$^2$ Institut Laue-Langevin, BP 156, 38042 Grenoble Cedex 9, France}

\ead{a.s.wills@ucl.ac.uk}

\begin{abstract}

The electric-field dependence of the magnetic structure of multiferroic MnWO$_4$ has been investigated using spherical neutron polarimetry as a function of temperature.  The application of an electric field drives the magnetic structure to a single `chiral' domain.  The magnitude of the field required to do this is shown to increase with temperature.  The link between the magnetic order and electric polarization is further demonstrated.  
\end{abstract}

\section{Introduction}

The magneto-electric effect, whereby the application of a magnetic field induces electric polarization and  the application of an electric field induces magnetization, has generated enormous interest lately due to its many possible applications \cite{NAH2000, WE2006}.  The recent revival is partly due to the identification of materials that become ferroelectric and ferromagnetic simultaneously. The frustrated, cycloidal, magnetic structure of these strongly coupled magneto-electrics gives rise to inverse Dzyaloshinsky-Moriya (DM) interactions \cite{TM1960, ID1958, IAS2006}, which, in turn, gives rise to an incommensurate modulation of the atomic structure with a propagation vector that is twice that of the magnetic propagation vector $\mathbf{k_l}$=2$\mathbf{k_m}$ \cite{KT2008}.  The ferroelectric polarization  $\mathbf{P_e}$ is proportional to the phase difference, $\phi$, between the chains of magnetically ordered ions \mbox{$\mathbf{P_e}\propto M_1 M_2 sin\phi [\mathbf{k}\times[\mathbf{e_1} \times \mathbf{e_2}]]$} where $M_1$ and $M_2$ are the magnitudes of the moments in the orthogonal directions $\mathbf{e_1}$ and $\mathbf{e_2}$ \cite{ZD2008, MM2006}.  The coupling shows great sensitivity to applied magnetic fields and, as we demonstrate in MnWO$_4$, to the application of an electric field \cite{KT2006, YY2008}.  Furthermore, the magnitude of the electric field that was required to effect a change in the magnetic order is shown to be highly sensitive to the temperature.

MnWO$_4$ belongs to the family of multiferroic materials \cite{KT2006, OH2006, AHA2006} with a frustrated, cycloidal magnetic order in the multiferroic phase.  MnWO$_4$ crystallizes in the wolframite structure with alternate layers of  manganese and tungsten perpendicular to the $a$ axis.  The structure is best described by the monoclinic $P2/c$ space group, with $\beta=91.1\,^{\circ}$ and lattice parameters $a = 4.82$\AA, $b = 5.75$\AA~and $c = 4.99$\AA.  The manganese is located at the 2$f$ site and forms zig-zag chains of Mn$^{2+}$$(S=5/2)$ ions surrounded by edge sharing oxygen octahedra.  MnWO$_4$ has been shown to undergo three successive antiferromagnetic phase transitions that generate long wavelength magnetic structures \cite{GL1993, HE1997}.  The first magnetically ordered state, AF3, occurs below $T_N$ = 13.5 K and is sinusoidally modulated with the easy direction in the $ac$ plane forming an angle of $35^{\circ}$ to the $a$ axis.  The second order transition to the AF2 phase at $T_{2}= 12.5$K introduces a component in the [010] direction that gives rise to the cycloidal magnetic order, with incommensurate (ICM) propagation vector  \mbox{$\mathbf{k_2}$= (-0.214, 0.5, 0.457)}.  The final, first order, transition into the commensurate (CM) AF1 phase occurs at $T_{1}\approx7$K and the structure becomes sinusoidally modulated once more with propagation vector  $\mathbf{k_1}$= (±$\frac{1}{4}$, $\frac{1}{2}$, $\frac{1}{2}$) \cite{GL1993}.  The AF2 phase has been demonstrated to be multiferroic, with the $\mathbf{P_e}$ along with the $b$ axis, perpendicular to the easy plane  \cite{AHA2006} . 

The magnetic structure was determined using spherical neutron polarimetry (SNP), the technique that gives the most accurate measurement of  the `chiral' terms of the magnetic tensor to give an insight into the phase and chirality of magnetic structures.\cite{PJB2005}.

\section{Spherical Neutron Polarimetry}

\begin{figure}
\centering
\begin{minipage}[t]{0.55\linewidth}
\centering
\includegraphics[scale=0.4]{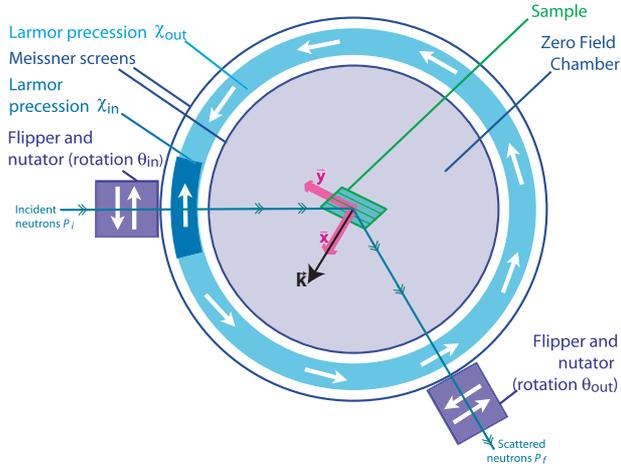}
\end{minipage}
\begin{minipage}[b]{0.4\linewidth}
\caption{(Color online) Schematic diagram of the CRYOPAD (CRYOgenic Polarization Analysis Device).  The sample is placed in a `zero field chamber' with incoming neutrons, $P_i$, oriented arbitrarily in the orthogonal axis system $\mathbf{x}$ marked in relation to $\mathbf{Q}$, $\mathbf{z}$ is with the zone axis and out the page and $\mathbf{y}$ makes up the right hand set.  The orthogonal components, $\mathbf{x}$, $\mathbf{y}$ and $\mathbf{z}$ of the scattered neutron beam,  $P_f$, are measured to give the $3\times3$ polarization matrix.}\label{cryo}
\end{minipage}
\end{figure}

\begin{figure}
\centering
\includegraphics[scale=0.35]{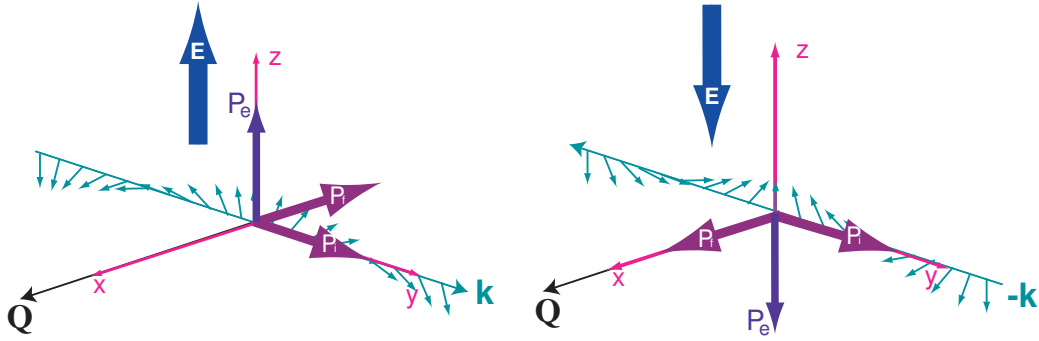}
\caption{Idealized orientation, with the propagation vector $\mathbf{k_1}$lying fully in the scattering plane with the sine component and electric  polarization $\mathbf{P_e}$ parallel to $\mathbf{z}$  and the cosine component with $\mathbf{y}$.  The applied electric field is along $\mathbf{E}$ parallel to $\mathbf{P_e}$.  The scattering vector is denoted $\mathbf{Q}$.  The polarization axes are marked  $\mathbf{x}$,  $\mathbf{y}$  and $\mathbf{z}$ and the direction of polarization of the incoming beam is with $P_i$ and the final polarization, due to the `chiral' scattering is with $P_f$.}
\label{axes}
\end{figure}

The SNP technique allows the polarized moments of the
neutron beam to be arbitrarily oriented using a combination of two
rotations corresponding to the spherical coordinate system
\cite{ELB2005, AP2007}.  All components of the
polarization matrix, with polarization axes  defined with $\mathbf{x}$
parallel to the scattering vector $\mathbf{Q}$, $\mathbf{z}$ with the zone axis of the crystal and $\mathbf{y}$ completing the right handed cartesian
set, are recorded, figure \ref{cryo}.  The polarization matrix is the experimental
quantity that is most closely related to the polarization tensor
form of the Blume Maleev equations, fully described in \cite{MB1963, PJB1993}.   The perpendicular sine and cosine components of the cycloidal magnetic structure, when they lie perpendicular to the $\mathbf{Q}$,  give rise to a non-zero `chiral term' as polarization is generated in the $\mathbf{x}$ direction, described by the cross product term in the Blume Maleev equations, \mbox{$\textbf{P}_{f_{chiral}}= -i(\textbf{M}_{\perp}^{*} \times \textbf{M}_{\perp})$}.   An idealized experimental arrangement, which would have both components perpendicular to $\mathbf{Q}$, is shown in figure \ref{axes}.  In this instance the polarization tensor will have the simplified form:

\begin{equation}
\begin{array}{ccc}
\mathcal{P}= \left(\begin{array}{ccc}
-1  &  0  &  0 \\
B &  A  &  0\\
B &  0  &  -A \\
 \end{array}\right)&
 \textrm{     or    }&
 
 \mathcal{P}= \left(\begin{array}{ccc}
-1  &  0  &  0 \\
-B &  A  &  0\\
-B &  0  &  -A \\
 \end{array}\right)
\end{array}
\end{equation}

\noindent
With the terms,

\begin{equation}
\begin{array}{ccc}

A=\frac{M_1(\textbf{k})^2-M_2(\textbf{k})^2}{M_1(\textbf{k})^2+M_2(\textbf{k})^2} &

\textrm{and}&

B=\frac{2M_1(\textbf{k})^2M_2(\textbf{k})^2}{M_1(\textbf{k})^2+M_2(\textbf{k})^2}
 
 \end{array}
\end{equation}

The $A$ component of the matrix gives the ellipticity of the cycloid, it is unity when the structure is spherical and the sine and cosine components have equivalent magnitudes. The $B$ component is the chiral term of the matrix, the magnitude is dependent on the population of each of the `chiral' domains, $\pm$ $\mathbf{k}$; when $\pm$ $\mathbf{k}$ are equally populated $B$=0; when the magnetism is ordered as a single $\mathbf{k}$ domain $B$ will have a magnitude of one, with the sign of $B$ dependent on the handedness of $\mathbf{k}$, shown in figure \ref{axes}.  

As stated previously this is an idealised experimental arrangement, however, the matrix element $B$ is always non-zero when there is a real and imaginary part of the magnetic interaction vector perpendicular to $Q$.

\section{Results and Discussion}

The polarization matrix was measured using CRYOPAD on D3 at the ILL.  The crystal was oriented with the [111] direction as the zone axis, along the $\mathbf{z}$ direction.   The polarization of the reflection \mbox{$(1\bar{1}\bar{1})$ - $\mathbf{k_2}$} was measured to follow the changes in the $B$ component.  The electric field was applied along the $b$ axis of the single crystal via sputtered gold contacts \cite{APunpub}.  The field was ramped from \mbox{0Vmm$^{-1}$} to \mbox{180Vmm$^{-1}$} to \mbox{0Vmm$^{-1}$} with constant temperature and then the polarity of the field was reversed and the process repeated.  The crystals were cooled to the constant temperatures, 12K, 11K and 10K with a poling electric field applied.

\begin{figure}
\centering
\begin{minipage}[t]{0.4\linewidth}
\centering
\includegraphics[scale=0.24]{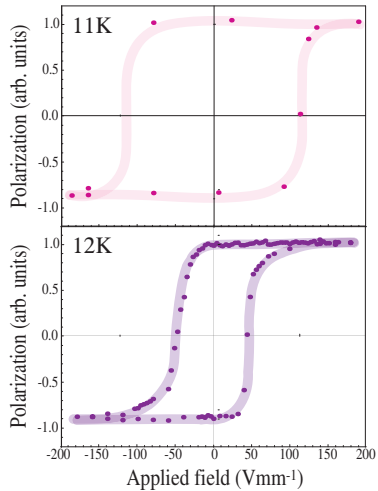}
\end{minipage}
\hspace{0.75cm}
\begin{minipage}[b]{0.45\linewidth}
\caption{Hysteresis of magnetic domain population determined by measurement of the chiral term of the polarization matrices recorded at 11K (top) and 12K (bottom).  The coercive field can be seen to be \mbox{$\approx$100Vmm$^{-1}$} and \mbox{$\approx$50Vmm$^{-1}$} respectively.  The polarization does not achieve the arbitrary value of  -1 due to the magnetic reflection not being  completely aligned in the plane. Thus, some of the information that would be transmitted in the $\mathbf{x}$  channel of the polarization matrix is transmitted in the $\mathbf{z}$. The lines are a  guide for the eyes.}\label{loop}
\end{minipage}
\end{figure}

The coercive field at 12K is \mbox{$\approx$50Vmm$^{-1}$} a reduction in temperature by 1K to 11K increased the coercive field to \mbox{$\approx$100Vmm$^{-1}$}.  At 10K the coercive field could not be achieved with \mbox{180Vmm$^{-1}$}, but a reduction in the spin polarization was observed, an indication that at \mbox{$\approx$200Vmm$^{-1}$} the sign of $B$ would be inverted. The complete hysteresis loops recorded can be seen in figure \ref{loop}.  

The experiment was repeated at 13K in the higher temperature, ICM phase AF3 and at 6K in the lower temperature CM AF1 phase.  There was no significant contribution to the `chiral' terms of the polarization matrix at either of these temperatures and no hysteresis was recorded.

Overall, the sensitivity of the coercive field to the temperature is surprisingly large, approximately doubling as the temperature is reduced by 1K.   The point at which the `chiral' term becomes zero implies the $\pm$$\mathbf{k_2}$ domains are indistinguishable.  The coercive field can be considered to be the field which is required to drive one of the chiral domains to have the same population equally populated, the $+\mathbf{k_2}$ and the $-\mathbf{k_2}$ domains are equivalent.  This interpretation leads to the canceling of one of the components and the loops may be understood as the inter-conversion of the $+\mathbf{k_2}$ and the $-\mathbf{k_2}$ domains.  This is possible as a cycloidal structure is not truly chiral, a $\pi$ rotation about the axis of $\mathbf{k}$ followed by a phase shift of $\pi/2$ along $\mathbf{k}$ will generate the oppositely `handed' domain.  Thus, the coercive field can be considered as the point at which the $\pi$ rotation about $\mathbf{k}$ occurs, and the saturation as where the spin chain translates to optimize $\phi$.

\section{Conclusion}

MnWO$_{4}$  shows the spin polarization indicative of a magnetic structure with perpendicular sine and cosine components.  The magnetic domain population has been controlled by the application of an electric field showing the link between magnetism and ferroelectricity, with a large sensitivity to temperature, in the multiferroic phase.  The electric field had no influence on the magnetic structures in the AF3 or the AF1 phases showing that there is no magneto-electric coupling.

\section{Acknowledgments}
AP would like to thank the EPSRC (grant number EP/D053560) for financial support and A. T. Boothroyd and D. Prabhakaran for supplying the sample and Z. Davies for useful discussions.

\section*{References}

\providecommand{\newblock}{}


\begin{thebibliography}{10}
\expandafter\ifx\csname url\endcsname\relax
  \def\url#1{{\tt #1}}\fi
\expandafter\ifx\csname urlprefix\endcsname\relax\def\urlprefix{URL }\fi
\providecommand{\eprint}[2][]{\url{#2}}

\bibitem{NAH2000}
Hill N~A 2000 {\em J. Phys. Chem. B\/}  6694

\bibitem{WE2006}
Eerenstein W, Mathur N~D and Scott J~F 2006 {\em Nature\/}  759

\bibitem{TM1960}
Moriya T 1960 {\em Phys. Rev.\/}  91

\bibitem{ID1958}
Dzyaloshinsky I 1957 {\em J. Phys. Chem. Solids\/}  241

\bibitem{IAS2006}
Sergienko I~A and Dagotto E 2006 {\em Phys. Rev. B\/}  094434

\bibitem{KT2008}
Taniguchi K, Abe N, Sagayama H \textit{et al} 2008
  {\em Phys. Rev. B\/}  64408

\bibitem{ZD2008}
Davies Z, Poole A and Wills A~S {\em unpublished work\/}

\bibitem{MM2006}
Mostovoy M 2006 {\em Phys. Rev. Lett.\/}  067601

\bibitem{KT2006}
Taniguchi K, Abe N, Ohtani S \textit{et al} 2006 {\em Phys.
  Rev. Lett.\/}  097203

\bibitem{YY2008}
Yamasaki Y, Sagayama H, Goto T \textit{et al} 2008
  {\em Phys. Rev. Lett.\/}  219902

\bibitem{PJB2005}
Brown P~J, Forsyth J~B and Tasset F 2005 {\em Solid State Science\/}  682

\bibitem{OH2006}
Heyer O, Hollmann N, Klassen I \textit{et al} 2006 {\em J. Phys.: Condens. Matter\/}  L471

\bibitem{AHA2006}
Arkenbout A~H, Palstra T~T~M, Siegrist T and Kimura T 2006 {\em Phys. Rev. B\/}
   184431

\bibitem{GL1993}
Lautenschl{\"a}ger G, Weitzel H, Vogt T \textit{et al} 1993 {\em Phys. Rev. B\/}  6087

\bibitem{HE1997}
Ehrenberg H, Weitzel H, Heid C \textit{et al} 1997 {\em J. Phys.: Condens. Matter\/}  3189

\bibitem{ELB2005}
Leli{\`e}vre-Berna E, Bourgeat-Lami E, Fouilloux P, \textit{et al} 2005 {\em Physica B\/}  131

\bibitem{AP2007}
Poole A, Wills A~S and Leli{\`e}vre-Berna E 2007 {\em J. Phys.: Condens.
  Matter\/}  452201

\bibitem{MB1963}
Blume M 1963 {\em Phys. Rev.\/}  1670

\bibitem{PJB1993}
Brown P~J, Forsyth J~B and Tasset F 1993 {\em Proc. R. Soc. Lond. A\/}
  147

\bibitem{APunpub}
Poole A, Wills A~S, Boothroyd A \textit{et al} {\em unpublished work\/}

\end{thebibliography}
\end{document}